\documentclass[onecolumn,floatfix,superscriptaddress,a4paper,
               showpacs,showkeys,nofootinbib,preprint]{revtex4}
\usepackage{epsfig}
\usepackage{latexsym}
\usepackage{xspace}
\usepackage{hyperref}
\usepackage[latin2]{inputenc}
\usepackage{indentfirst}
\usepackage{enumerate}

\usepackage{amsmath}
\usepackage{amssymb}
\usepackage[english]{babel}
\usepackage{url}

\newcommand{\eq}[1]{\begin{align} #1 \end{align}}
\newcommand{\bs}{\boldsymbol}

\begin{document}

\title{Van der Waals Equation of State with Fermi Statistics for\\
Nuclear Matter}
\author{V. Vovchenko}
\affiliation{
Taras Shevchenko National University of Kiev, 03022 Kiev, Ukraine}
\affiliation{
Frankfurt Institute for Advanced Studies, Johann Wolfgang Goethe University, D-60438 Frankfurt, Germany}
\affiliation{
GSI Helmholtzzentrum f\"ur Schwerionenforschung GmbH, D-64291 Darmstadt, Germany}
\author{D. V. Anchishkin}
\affiliation{
Bogolyubov Institute for Theoretical Physics, 03680 Kiev, Ukraine}
\affiliation{
Taras Shevchenko National University of Kiev, 03022 Kiev, Ukraine}
\affiliation{
Frankfurt Institute for Advanced Studies, Johann Wolfgang Goethe University, D-60438 Frankfurt, Germany}
\author{M. I. Gorenstein}
\affiliation{
Bogolyubov Institute for Theoretical Physics, 03680 Kiev, Ukraine}
\affiliation{
Frankfurt Institute for Advanced Studies, Johann Wolfgang Goethe University, D-60438 Frankfurt, Germany}

\date{\today}


%%%%%%%%%%%%%%%%%%%%%%%%%%%%%%%%%%%%%%%%%%%%%%%%%%%%%%%%%%%%%%%%%%%%%%%%%%%%%%%%
\begin{abstract}
The van der Waals (VDW) equation of state is
a simple and popular
model to describe  the pressure function in
equilibrium systems of particles with
both repulsive and attractive interactions.
This equation predicts an existence of a first-order
liquid-gas phase transition and
contains a
critical point.
Two steps to extend
the VDW equation  and make it appropriate for new physical applications
are carried out in this paper:
1) the grand canonical ensemble
formulation; 2) the inclusion of the quantum statistics.
The VDW equation with Fermi statistics is then applied
to a description of
the system of
interacting nucleons.
The VDW  parameters $a$ and $b$ are fixed to
reproduce the properties of
nuclear matter at
saturation density $n_0=0.16$~fm$^{-3}$ and zero temperature.
The model predicts a location of the
critical point
for the symmetric nuclear matter
at temperature $T_c\cong 19.7$~MeV and
nucleon number density $n_c \cong 0.07$~fm$^{-3}$.
\end{abstract}
%%%%%%%%%%%%%%%%%%%%%%%%%%%%%%%%%%%%%%%%%%%%%%%%%%%%%%%%%%%%%%%%%%%%%%%%%%%%%%%%

\maketitle

\section{Introduction}
The van der Waals equation of state is
a simple
analytical model
of the pressure function $p$ for equilibrium systems of  particles with both attractive and
repulsive interactions. The VDW model contains the first-order liquid-gas  phase
transition which ends at
the critical point. In the canonical ensemble (CE), where
independent variables are temperature $T$, volume $V$,
and number of particles $N$, the VDW equation of state has
most simple and transparent form,
(see, e.g., Refs.~\cite{greiner,LL}),
\eq{\label{eq:vdw}
p(T,n) ~=~ \frac{NT}{V-bN} ~-~ a \frac{N^2}{V^2}~ \equiv~\frac{n\,T}{1-bn}~-~a\,n^2~,
}
where $a>0$ and $b>0$ are the VDW parameters which describe
attractive and repulsive interactions, respectively, and $n\equiv N/V$ is the particle number density.
The first term in the right-hand-side of Eq.~\eqref{eq:vdw} corresponds
to the excluded volume (EV) correction, which manifests
itself in a substitution of a total volume $V$
by the available volume, $V_{\rm av} = V - b\,N$. The second
term comes from the mean field which describes attractive interactions between particles.
In order to apply the VDW equation of state
to systems with variable number of particles it is necessary to
switch to the grand canonical ensemble (GCE).
This procedure was first performed for the
EV model, i.e., for $a=0$ in Eq.~(\ref{eq:vdw}),  in Refs.~\cite{vdw-1,vdw-2}.
In our recent paper \cite{VdW-GCE}, the full VDW equation \eqref{eq:vdw},
with both attractive and repulsive terms, was transformed
from the CE to the GCE for
systems with Boltzmann statistics.
There are several physical situations when the GCE formulation is desirable
(see Ref.~\cite{VdW-GCE} for details).
Note that the EV and VDW models
can also be conveniently treated
within the GCE
in a framework of the thermodynamic mean-field approach
(see Refs.~\cite{mf-1992,mf-1995,mf-2014}).

Equation \eqref{eq:vdw} is valid for classical systems, where
the effects of quantum statistics are neglected.
In the present paper we suggest a generalization
of the VDW equation to include effects of the quantum
statistics. Proper treatment of quantum effects
appears to be crucially important for a description
of statistical equilibrium at small temperatures.
The quantum statistics formulation is much easier
to introduce in the GCE than in the CE.
This is an additional physical example where the GCE formulation is particularly helpful.
Thus, we use our recent
results of the GCE formulation~\cite{VdW-GCE}
as a starting point for a quantum generalization of the VDW equation of state.
As a next step, the VDW equation of state with Fermi statistics is used to describe
nuclear matter. The VDW parameters, $a$ and $b$, which correspond, respectively, to attractive and
repulsive interactions between nucleons, are fixed
to reproduce the properties of the symmetric nuclear matter at
zero temperature: saturation density $n_0=0.16$~fm$^{-3}$, binding energy per nucleon,
$- 16$~MeV; and zero pressure, $p=0$.

The paper is organized as follows.  In Sec.~\ref{sec-vdw} the VDW
equation of state is transformed
into the GCE, and the quantum statistical formulation of this equation is elaborated
in Sec.~\ref{sec:vdw-q}.
In Sec.~\ref{sec-nm} the VDW equation of state with Fermi statistics
is applied to a description of nuclear matter.
A summary in Sec.~\ref{sec:summary}
closes the article.

\section{VDW equation for the Boltzmann statistics in the GCE}
\label{sec-vdw}
The VDW pressure function (\ref{eq:vdw}) corresponds to the
Boltzmann approximation, i.e., the effects of quantum statistics (Bose or Fermi)
are neglected. In our recent paper~\cite{VdW-GCE} the VDW equation of state
was formulated in the GCE.  The GCE
pressure, $p(T,\mu)$, is a function of temperature $T$ and chemical potential $\mu$.
It contains a complete information about thermodynamical functions of the system.
Particle number density $n(T,\mu)$, entropy density $s(T,\mu)$,
and energy density $\varepsilon(T,\mu)$ can be  presented
in terms of $p$ and its $T$ and $\mu$ derivatives:
\eq{\label{pTmu-s}
n(T,\mu)=\left(\frac{\partial p}{\partial \mu}\right)_T~,~~~~
s(T,\mu)=\left(\frac{\partial p}{\partial T}\right)_{\mu}~,~~~~\varepsilon (T,\mu)=
T \left(\frac{\partial p}{\partial T}\right)_{\mu}+
\mu \left(\frac{\partial p}{\partial \mu}\right)_{T}-p~.
}
For $a=b=0$ the above VDW equations are reduced to the ideal gas expressions
for classical particles.

The VDW equation of state in the GCE is obtained in the
form of a transcendental equation for particle number density
$n \equiv n(T,\mu)$ as a function of
$T$ and $\mu$ \cite{VdW-GCE}:
\eq{\label{eq:nvdwtr}
n(T,\mu) ~& =~ \frac{n^{\rm id}(T, \mu^*)}{1~+~b \,n^{\rm id}(T, \mu^*)}~,~~~~~
\mu^* ~=~ \mu~ -~ T \frac{bn}{1\,-\,bn} ~+~ 2 a n~,
}
where $n^{\rm id}$ is a particle number density in the ideal  Boltzmann gas,
\eq{
n^{\rm id}(T, \mu) ~ &=~
\exp\left(\frac{\mu}{T}\right)\,\frac{d\,m^2\, T}{2 \pi^2} \, K_2\left(\frac{m}{T}\right)~,\label{n-id}
}
with $d$ being the degeneracy factor and $m$ the particle mass, $K_2(x)$ the Bessel function.
Note that the relativistic form of a dispersion relation is considered,
$\omega(k)=\sqrt{m^2+k^2}$, where $\omega$ and $k$ are the free particle energy and momentum,
respectively. The GCE VDW pressure $p(T,\mu)$ is then obtained
by inserting $n(T,\mu)$ (\ref{eq:nvdwtr}) into Eq.~(\ref{eq:vdw}).

The VDW pressure (\ref{eq:vdw}) is a unique function of variables $T$ and $n$
for all $T\ge 0$ and $0\le n\le 1/b$.
The VDW equation
of state contains
a first-order liquid-gas phase transition and has a critical point. The critical point
$(T_c,n_c)$ corresponds to the temperature and particle number density, where
\eq{\label{p-der}
\left(\frac {\partial p}{\partial n}\right)_T~=0, \quad\left(\frac {\partial ^2 p}{\partial n^2}\right)_T~=~0~.
}
The
thermodynamical quantities at the critical point
are equal to
\cite{greiner,LL}:
\eq{\label{crit}
 T_c = \frac{8a}{27b}~,~~~~~ n_c =
\frac{1}{3b}~,~~~~~ p_c = \frac{a}{27b^2}~.
}
At $T>T_c$ the following equation is  always valid,
\eq{\label{p-der-1}
\left(\frac {\partial p}{\partial n}\right)_T~>~0~,
}
while at $T<T_c$ the unstable region appears with
\eq{\label{p-der-2}
\left(\frac {\partial p}{\partial n}\right)_T~<~0~.
}
Therefore, the VDW isotherm $p(n,T)$ at $T<T_c$ has a local maximum at $n=n_1$ and a local minimum
at $n=n_2>n_1$. The unstable part (\ref{p-der-2}) of the VDW isotherm at the interval
$[n_1,n_2]$, together
with two additional parts -- $[n_g,n_1]$ and $[n_2,n_l]$ (they are called metastable) --
are transformed to a mixture of two phases: a gas with density $n_g<n_1$
and a liquid with density $n_l>n_2$. This is done according to the Maxwell rule
of the equal areas (see, e.g., Refs.~\cite{greiner,LL}), which leads to a constant pressure $p(T,n_g)=p(T,n_l)$
inside the density interval $[n_g,n_l]$.

In the GCE the mixed phase region appears in a different way.
At $T>T_c$ there is a unique solution of Eq.~\eqref{eq:nvdwtr}, while
at $T<T_c$ it
may have either one solution or three different solutions for particle number density
$n(T,\mu)$.
Therefore, either one or three different solutions
may also appear for the VDW pressure $p(T,\mu)$.
In a case when three different values of $p(T,\mu)$ are possible,
the solution with a largest pressure survives
in accordance to the Gibbs criterion (see Appendix~\ref{app:GibbsMaxwell} for details).
The gas-liquid mixed phase in the $T$-$\mu$ plane belongs to the
line $\mu=\mu(T)$, where the solutions  $n_g(T,\mu)$ and $n_l(T,\mu)$
correspond to equal pressures,
$p_g(T,\mu)=p_l(T,\mu)$.

The classical Boltzmann statistics leads to nonphysical behavior in zero temperature limit.
This is already seen on an ideal gas level.
For the ideal Boltzmann gas an entropy density in non-relativistic limit $T/m\ll 1 $
is equal to
\eq{\label{s-boltz}
s^{\rm id}_{\rm Boltz}~\cong~\frac{n^{\rm id}}{T}\,\left[ m~+~\frac{5}{2}T~-\mu\right]~.
}
In Eq.~(\ref{s-boltz}) we use expressions $p^{\rm id}=n^{\rm id}T$ and $\varepsilon^{\rm id}\cong
n^{\rm id}(m+3T/2)$ for the ideal gas pressure and (non-relativistic) energy density, respectively.
Using an asymptotic expansion for the $K_2$ Bessel function at large arguments, $K_2(x)\cong \sqrt{\pi/(2x)}\exp(-x)$,
one finds from Eq.~(\ref{n-id}) that to have a finite (nonzero) value $n_0$ of particle number density at $T\rightarrow 0$
the chemical potential should be equal to
\eq{\label{mu-0}
\mu~\cong~m~-~\frac{3\,T}{2}\,\ln(T/c_0)~,~~~~~~~ c_0~=~\frac{2\pi\,n_0^{2/3}}{m}~.
}
Thus, only one limiting value, $\mu=m$, is admitted in the Boltzmann gas at $T=0$ (this corresponds to
zero value of chemical potential, $\mu_{\rm non-rel}\equiv \mu-m$, used
in non-relativistic statistical physics). For $\mu>m$ or $\mu<m$ at $T=0$ one finds
for the particle number density $n=0$ or $n=\infty$, respectively.

Therefore, the entropy of the ideal Boltzmann gas (\ref{s-boltz}) at $T\rightarrow 0$ is
\eq{\label{s-id-T0}
s^{\rm id}~\cong~n_0\,\left[\frac{5}{2}~+~\frac{3}{2}\ln(T/c_0)\right]~,
}
and it becomes negative in zero temperature limit, in a contradiction
with the 3${\rm rd}$ law of thermodynamics.
The quantum statistics is needed to describe a physical
system at $T\rightarrow 0$.

%%%%%%%%%%%%%%%%%%%%%%%%%%%%%%%%%%%%%%%%%%%%%%%%%%%%%%%
\section{VDW equation of state with quantum statistics}
\label{sec:vdw-q}
%%%%%%%%%%%%%%%%%%%%%%%%%%%%%%%%%%%%%%%%%%%%%%%%%%%%%%%

Quantum generalization of the VDW equation of state is not a trivial task.
Let us outline some general requirements for the quantum version of this
equation of state:
\begin{enumerate}
\item It should be transformed to the ideal {\it quantum} gas
at $a=0$ and $b=0$.
\item It should be equivalent
to the classical VDW equation of state \eqref{eq:vdw}
in a region of thermodynamical parameters where quantum statistics can be neglected.
\item The entropy should be a non-negative quantity and go to zero at $T\rightarrow 0$.
\end{enumerate}

The pressure of the ideal quantum gas in the GCE reads
\begin{equation}
p^{\rm id} (T, \mu) ~ = ~\frac{d}{3} \int \frac{d^3k}{(2\pi)^3} \,
\frac{\bs k^2}{\sqrt{m^2+\bs k^2}} \, \left[ \exp\left(\frac{\sqrt{m^2+\bs k^2}-\mu}{T}\right)
+ \eta\right]^{-1}~.
\label{p-id}
\end{equation}
In Eq.~(\ref{p-id}), $\eta$ equals +1 for Fermi statistics, -1 for Bose statistics,
and 0 for the Boltzmann approximation.
All other thermodynamical functions can be calculated from Eqs.~(\ref{pTmu-s}).
Ideal quantum gas expressions for thermodynamical functions satisfy
the 3rd law of thermodynamics, i.e., $s\ge 0$ and $s\rightarrow 0$ at $T\rightarrow 0$.

Let us now formulate a generalization of the VDW equation of state which
includes effects of the quantum statistics.
Note that $p(T,\mu)$ for the Boltzmann case can be rewritten using
Eq.~\eqref{eq:nvdwtr} as
\begin{equation}
p(T,\mu)~ = ~p^{\rm id} (T, \mu^*) - a\,n^2~,
\label{eq:pq}
\end{equation}
where
\begin{equation}
\mu^*~ =~ \mu~ - ~b \, p(T,\mu) - a\,b\,n^2 + 2 \, a \, n~.
\label{eq:pq-2}
\end{equation}
The function $p^{\rm id}$ in Eq.~(\ref{eq:pq}) corresponds to the ideal gas pressure in
the Boltzmann approximation, i.e. $\eta=0$ in Eq.~(\ref{p-id}).
We suggest the quantum VDW equation of state in the same form as Eq.~\eqref{eq:pq}
but with ideal {\it quantum} gas pressure  $p^{\rm id}$, i.e., for quantum case we propose to
take $\eta=\pm 1$ in Eq.~(\ref{p-id}), which corresponds to the Fermi or Bose
statistics.

In accordance with (\ref{pTmu-s}), one has the following for the particle number density:
\begin{equation}
n(T,\mu) ~\equiv ~ \left(\frac{\partial p}{\partial \mu}\right)_T~
=~n^{\rm id} (T, \mu^*) \, (1-bn) \, \left(1 + 2 \, a\,
\frac{\partial n}{\partial \mu} \right) - 2\, a \, n\, \frac{\partial n}{\partial \mu}\,.
\end{equation}
This equation can be transformed to
\begin{equation}
\Big[ n^{\rm id} (T, \mu^*) \, (1-bn) - n \Big]
\left(1 + 2\, a \, \frac{\partial n}{\partial \mu}\right)\, =\, 0~.
\label{eq:eqnq}
\end{equation}
The solution of this equation, which has a physical meaning, reads
\begin{equation}
n(T,\mu) = \frac{n^{\rm id}(T,\mu^*)}{1 + b  n^{\rm id}(T,\mu^*)}~,~~~~
n^{\rm id} (T, \mu)  = \frac{d}{2\pi^2} \int_0^{\infty} dk k^2
\left[ \exp\left(\frac{\sqrt{m^2+k^2}-\mu}{T}\right) + \eta\right]^{-1},
\label{eq:nvdwq}
\end{equation}
and it has the same form as  Eq.~\eqref{eq:nvdwtr}.
However, a principal difference
is that $n^{\rm id}$ in Eq.~(\ref{eq:nvdwq}) is a particle number
density of the ideal {\it quantum} gas,
whereas $n^{\rm id}$ in Eq.~(\ref{eq:nvdwtr}) corresponds to the ideal
{\it classical} gas, i.e., $\eta=0$, and is given by Eq.~\eqref{n-id}.
We also note that, in the quantum case, expression (\ref{eq:pq-2}) for the shifted
chemical potential $\mu^*$ should be used instead of (\ref{eq:nvdwtr}).

Equations \eqref{eq:pq} and \eqref{eq:nvdwq} correspond to the system of two
equations for two unknown functions: $p(T,\mu)$ and $n(T,\mu)$.
The VDW model defined by these equations possesses all the required properties.
First, at $a=0$ and $b=0$ Eqs.~\eqref{eq:pq} and \eqref{eq:nvdwq} are reduced
to the ideal {\it quantum} gas expressions. Second, for those $T$ and $\mu^*$ values,
where {\it quantum} expressions for $p^{\rm id}$ and $n^{\rm id}$ can be approximated
by the Boltzmann statistics, i.e., by Eq.~\eqref{n-id} for $n^{\rm id}$ and $p^{\rm id}=Tn^{\rm id}$
for the ideal gas pressure, Eqs.~\eqref{eq:pq} and \eqref{eq:nvdwq}
become automatically equivalent
to the classical VDW equation of state \eqref{eq:vdw}.
Third, the entropy density has the following form
\eq{\label{eq:svdwq}
s(T,\mu) ~ \equiv~ \left(\frac{\partial p}{\partial T}\right)_{\mu}~
=~ \frac{s^{\rm id}(T,\mu^*)}{1+b\,n^{\rm id}(T,\mu^*)}~,
}
thus, it is
always positive
for the {\it quantum} ideal gas expressions of $s^{\rm id}$,
and  $s \to 0$ at $T \to 0$.

The energy density can be calculated from Eq.~\eqref{pTmu-s} as
\eq{\label{eq:evdwq}
\varepsilon(T,\mu) ~=~ \Big[\overline{\epsilon}_{\rm id}(T, \mu^*)~-~a\,n\Big]\,n~,
}
where $\overline{\epsilon}_{\rm id}(T, \mu)$ is the
average energy per particle in the ideal gas,
\eq{\label{eq:epsilon}
\overline{\epsilon}_{\rm id}(T, \mu) ~
= ~\frac{\varepsilon^{\rm id}(T,\mu)}{n^{\rm id}(T,\mu)}~.
}

One can rewrite the VDW pressure as a function of temperature $T$ and particle
density $n$.
It follows from Eq.~\eqref{eq:nvdwq} that $\mu^*$ can be written as a function
of $T$ and $n$,
\begin{equation}
\mu^*(n,T) ~=~ \mu^{\rm id} \Big(\frac{n}{1-bn},T\Big)~,
\label{eq:mus-2}
\end{equation}
where $\mu^{\rm id} (n,T)$ is the chemical potential of the ideal quantum gas,
which is a solution of the following transcendental equation
for the given $n$ and $T$:
\begin{equation}
n ~=~ \frac{d}{2\pi^2} \int_0^{\infty} dk \, k^2 \,
\left[ \exp\left(\frac{\sqrt{m^2+k^2}-\mu^{\rm id}}{T}\right) + \eta\right]^{-1}~.
\end{equation}
Equation~\eqref{eq:pq} can be then rewritten as
\begin{equation}
p ~=~ p^{\rm id} \Big[T,\mu^{\rm id}\Big(\frac{n}{1-bn},T\Big)\Big] - a\,n^2~.
\label{eq:PnTq}
\end{equation}
One can easily check that this equation coincides with Eq.~\eqref{eq:vdw} in a
case of the Boltzmann statistics and, thus, can be indeed regarded as a quantum
generalization of the classical VDW equation in the CE.
It can be also instructive to consider a formulation of the VDW equation with quantum statistics within the thermodynamic mean-field approach developed in Refs.~\cite{mf-1992,mf-1995,mf-2014}. This is presented in Appendix~\ref{app:MF}.

%%%%%%%%%%%%%%%%%%%%%%%%
\section{Nuclear matter}
\label{sec-nm}
%%%%%%%%%%%%%%%%%%%%%%%%
In this section the VDW equation of state with quantum statistics is used
to describe the
properties of symmetric nuclear matter.
Namely, a Fermi gas of  nucleons ($m\cong 938$~MeV and $d=4$) is considered
with  attractive and repulsive interactions
described by the $a$ and $b$ VDW parameters, respectively.
A study of nuclear matter has a long history.
The thermodynamics of nuclear matter and its applications to 
the production of nuclear fragments in heavy ion collisions
were considered in Refs. \cite{JDGM,RMS,FaiRandrup,BBLZ,CsSt} in 1980s.
A review of these early developments can be found in Ref.~\cite{CsernaiKapusta}.
Nowadays, the properties of nuclear matter are described
by many different models,
particularly by those which employ a self-consistent mean field approach
~\cite{wal,ZM,Brockmann,Mueller:1996pm,SMFreview}.
Excluded-volume corrections in the mean-field models have been considered in
Refs.~\cite{vdw-2,mf-1995,NM-EV}.
%The models generally predict the existence of a liquid-gas phase transition with
%VDW-like isotherms. 
Experimentally, a presence of the liquid-gas phase transition
in nuclear matter was first reported in Refs.~\cite{Finn,Minich,Hirsch} 
by indirect observations. The first direct measurements of the nuclear caloric 
curve were done by the ALADIN collaboration~\cite{Pochodzalla},
later followed by other experiments~\cite{Natowitz,Karnaukhov}.

Our consideration will be restricted to small temperatures, $T\le 30$~MeV,
thus, a pion production will be neglected.
In the present work, we also neglect a possible formation of nucleon clusters (i.e., ordinary nuclei)
and baryonic resonances (like $N^*$ and $\Delta$), which may be important at low
and high baryonic density, respectively.
%{\bf
%Thus, in the present work, we do not consider the distribution of the nuclear fragments.
%}
%
Within these approximations, the number of nucleons $N$ becomes a conserved number
and an independent variable in the CE. The chemical potential $\mu$ of the GCE regulates
the number density of nucleons.
%

%%%%%%%%%%%%%%%%%%%%%%%%%%%%%%%%
\subsection{Properties at $T=0$}
For calculations of the thermodynamic functions in the GCE, Eqs. \eqref{eq:pq},
\eqref{eq:nvdwq}, \eqref{eq:svdwq}, and \eqref{eq:evdwq} will be used.
In terms of variable $\mu^*$ (\ref{eq:pq}), thermodynamical functions
of the quantum VDW gas can be presented in terms of the corresponding functions
of the  ideal quantum gas as the following:
\eq{\label{vdw-ne-id}
& n(T,\mu)=\frac{n^{\rm id}(T,\mu^*)}{1+b\,n^{\rm id}(T,\mu^*)}~,~~~~~
p(T,\mu)~=~p^{\rm id}(T,\mu^*)~-~a~\left[\frac{n^{\rm id}(T,\mu^*)}{1+b\,n^{\rm id}(T,\mu^*)}\right]^2~,\\
& \varepsilon(T,\mu)=\frac{\varepsilon^{\rm id}(T,\mu^*)}{1+b\,n^{\rm id}(T,\mu^*)}~
-~a~\left[\frac{n^{\rm id}(T,\mu^*)}{1+b\,n^{\rm id}(T,\mu^*)}\right]^2~,~~~~~
s(T,\mu)=\frac{s^{\rm id}(T,\mu^*)}{1+b\,n^{\rm id}(T,\mu^*)}~.
\label{vdw-sp-id}
}
At zero temperature the ideal gas quantities in Eqs.~(\ref{vdw-ne-id}) and (\ref{vdw-sp-id})
can be written as
\eq{
n^{\rm id} (T=0, \mu^*)~ & =~ \frac{d}{2\pi^2} \int\limits_0^{\sqrt{\mu^{*2}-m^2}} d k \, k^2 \,
=~ \frac{d}{6\pi^2} \, (\mu^{*2}-m^2)^{3/2}, \\
p^{\rm id} (T=0, \mu^*) ~& =~ \frac{d}{6 \pi^2} \int\limits_0^{\sqrt{\mu^{*2}-m^2}} d k \, \frac{k^4}{\sqrt{k^2+m^2}} \,=~
\frac{d}{48\pi^2} \, \mu^* \, \sqrt{\mu^{*2}-m^2} \, (2\mu^{*2}-5m^2)~\nonumber \\
&-~  \frac{d}{16 \pi^2} \, m^4 \, \ln \frac{m}{\mu^*+\sqrt{\mu^{*2}-m^2}}, \\
\varepsilon^{\rm id}(T=0, \mu^*)~ & ~= \frac{d}{2 \pi^2} \int\limits_0^{\sqrt{\mu^{*2}-m^2}} d k \, k^2 \, \sqrt{k^2+m^2}\,=~
\frac{d}{16\pi^2} \, \mu^* \, \sqrt{\mu^{*2}-m^2} \, (2\mu^{*2}-m^2)~ \nonumber \\
& + ~  \frac{d}{16 \pi^2} \, m^4 \, \ln \frac{m}{\mu^*+\sqrt{\mu^{*2}-m^2}}, \\
s^{\rm id} (T=0,\mu^*) ~ &  = ~ \lim_{T \to 0} \frac{\varepsilon^{\rm id}(T, \mu^*) 
+ p^{\rm id} (T, \mu^*) - \mu^* \, n^{\rm id} (T, \mu^*)}{T}~ =~ 0~.
}
%{\bf
%where it is implied that $\mu^* > m$. Note that, formally, there also exist solutions $n = 0$ and $p=0$ at $\mu^* = \mu < m$.
%}
%
We fix parameters $a$ and $b$ in such a way to reproduce
the properties of nuclear matter in its ground state (see, e.g., Ref.~\cite{norm-nm}),
i.e., it should be
$p=0$ and $\varepsilon / n = m + E_B \cong 922$~MeV at $T=0$ and
$n=n_0\cong 0.16$~fm$^{-3}$.
Here $E_B \cong - 16$~MeV is the binding energy per nucleon.
One then finds,  $a \cong 329$~MeV$\,$fm$^3$ and
$b \cong 3.42$~fm$^3$.
% ($r \cong 0.59$~fm).
%{\bf
Note that parameter $b$ of the proper particle volume
can be expressed in terms of hard-core radius $r$ as $b=16\pi r^3/3$.
This gives $r \cong 0.59$~fm for the hard-core nucleon radius.

\begin{figure}[ht]
\centering
\includegraphics[width=0.49\textwidth]{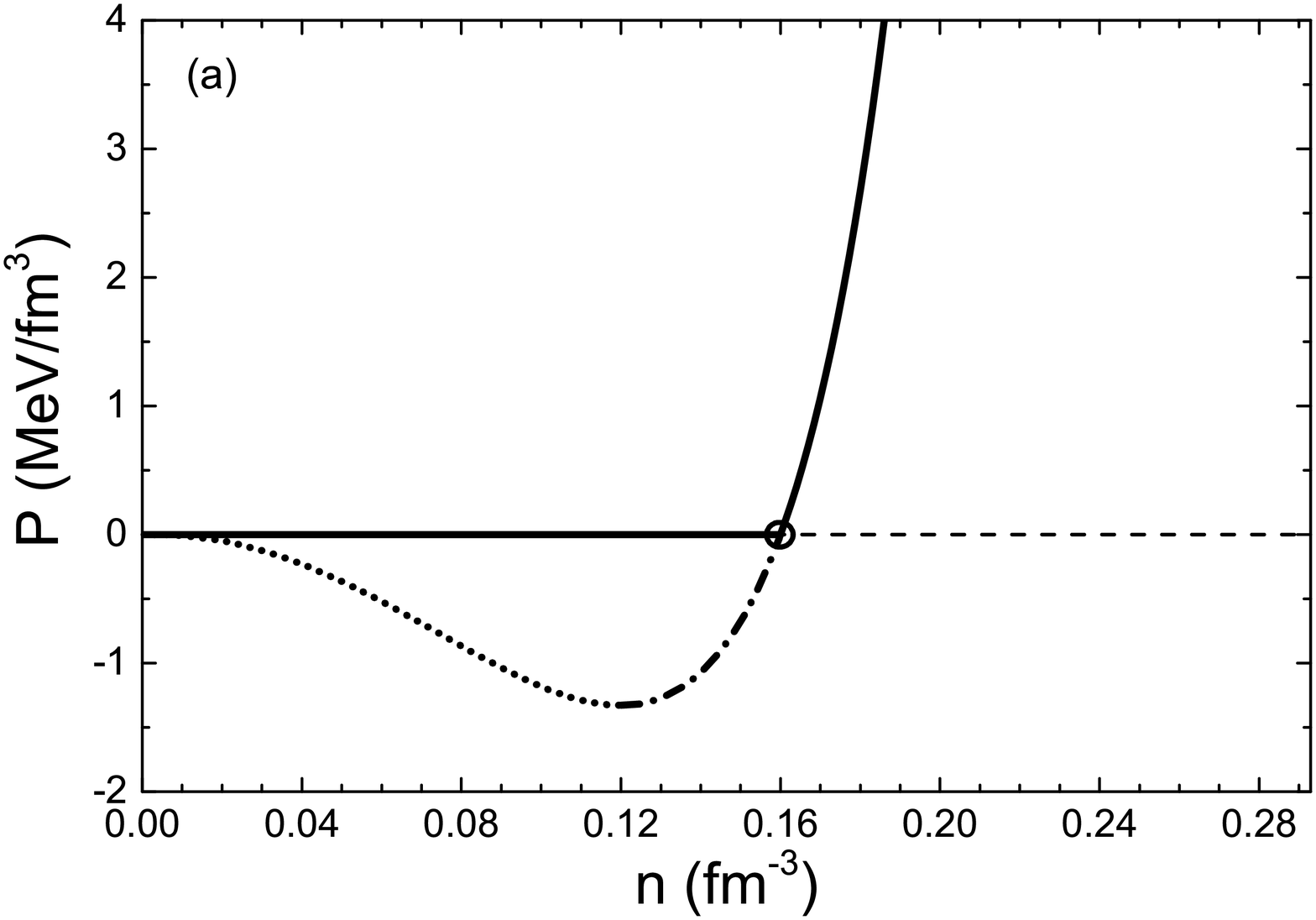}
\includegraphics[width=0.49\textwidth]{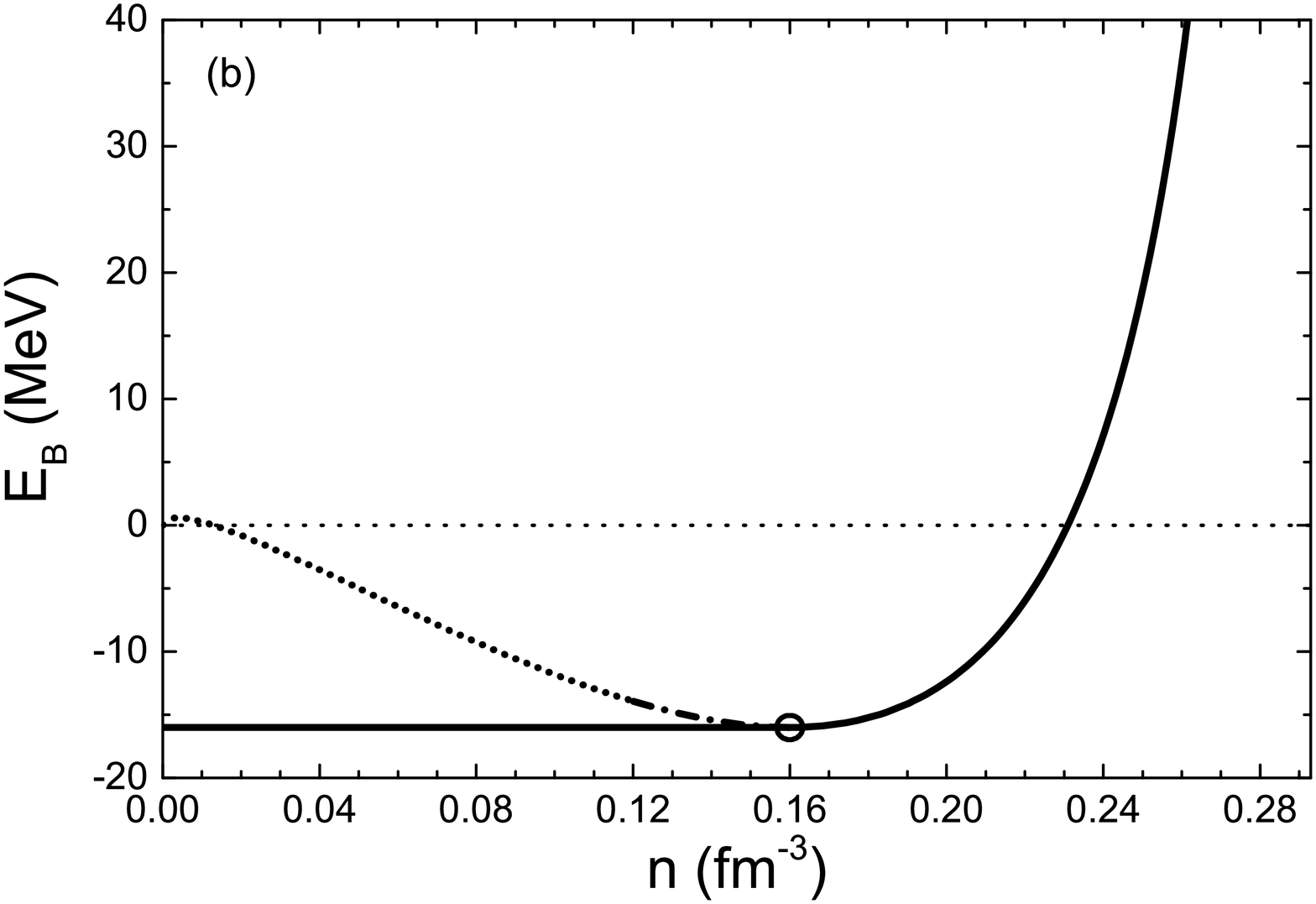}
\caption[]{Dependence of pressure $p$ (a)
and binding energy $E_B$ (b) on the
nucleon density $n$ at $T=0$. The VDW parameters are $a \cong 329$~MeV$\cdot$fm$^3$ and
$b \cong 3.42$~fm$^3$ ($r \cong 0.59$~fm).
The open circle corresponds to the ground state of nuclear matter.
The dash-dotted line corresponds to the metastable part
of VDW isotherm, whereas the dotted line corresponds to the unstable part.
}\label{fig-PE-T0}
\end{figure}

\begin{figure}[ht]
\centering
\includegraphics[width=0.49\textwidth]{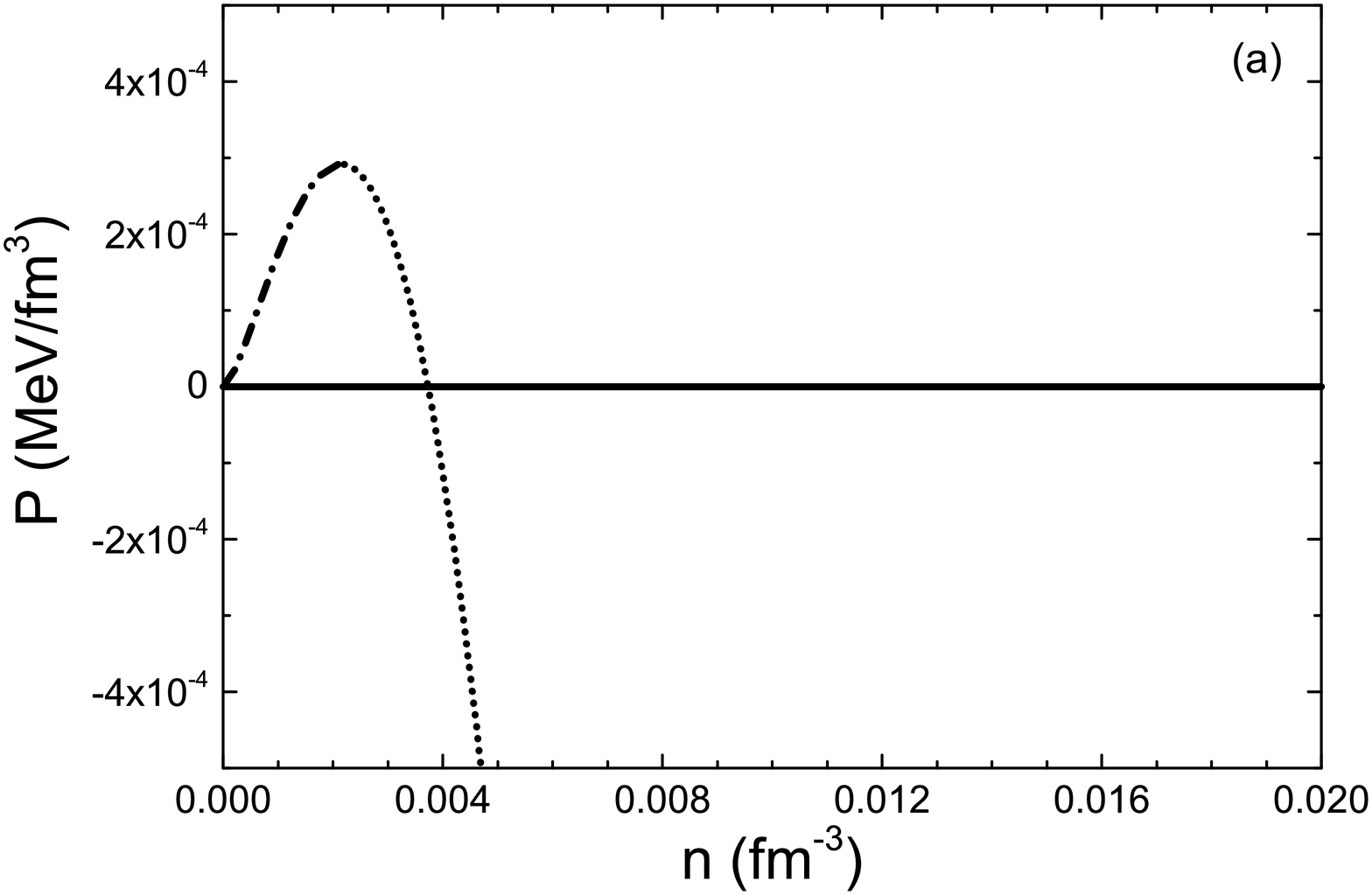}
\includegraphics[width=0.49\textwidth]{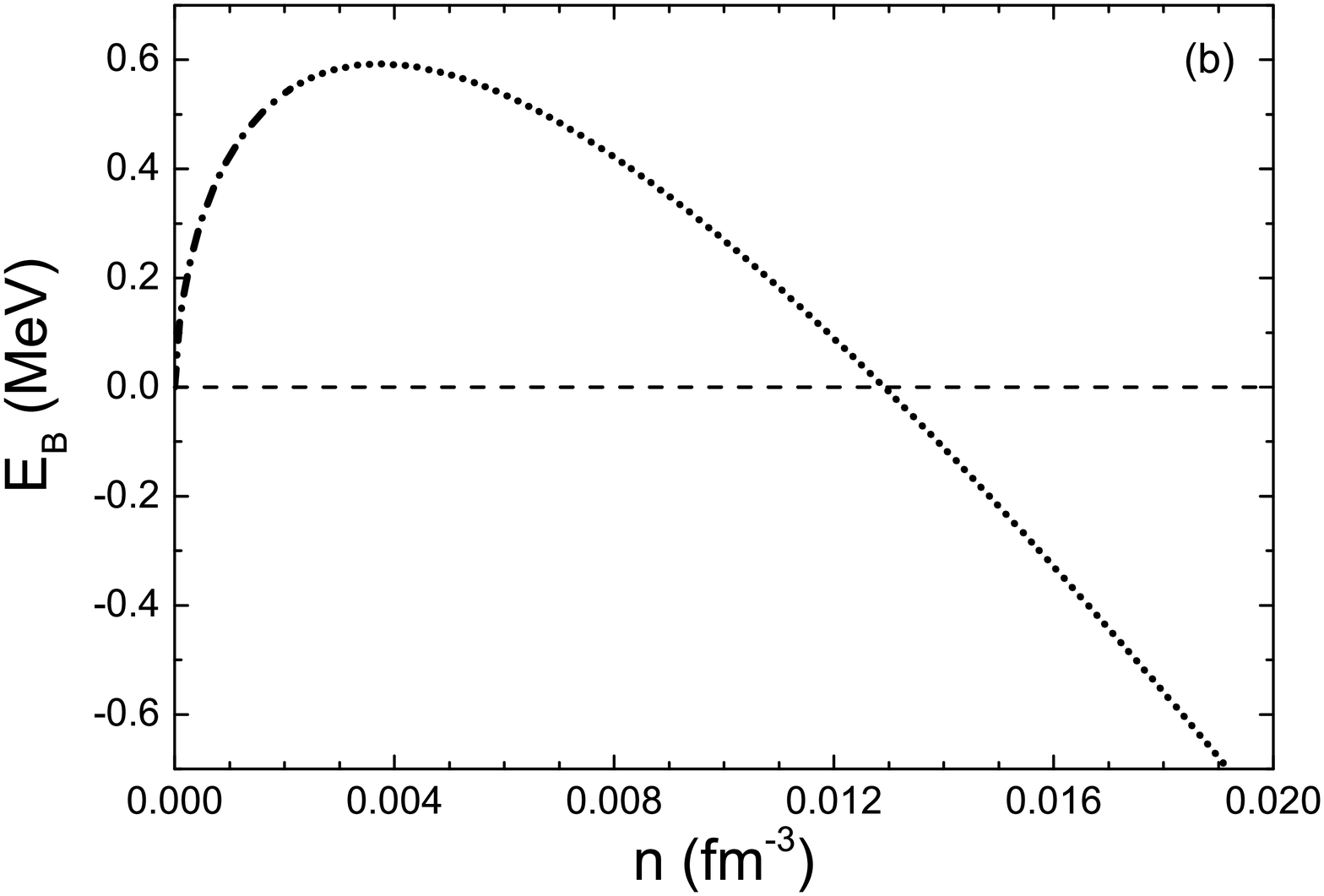}
\caption[]{
The same as in Fig.~\ref{fig-PE-T0}, but for
small values of $n$.
}\label{fig-PE-T0-small}
\end{figure}

The pressure $p$ and binding energy $E_B$ as
functions on  nucleon density $n$
at $T=0$ are shown in Fig.~\ref{fig-PE-T0} (a) and (b), respectively.
The stable VDW isotherms are depicted in Fig.~\ref{fig-PE-T0} by solid lines,
while the metastable and unstable parts are depicted by dash-dotted and dotted lines,
respectively. 
%We note that horizontal part of the isotherm corresponds
%to the mixed phase and results from the Maxwell construction~(see more details below).
%
At very small densities a gaseous  phase with almost ideal gas behavior is always present.
At $T=0$ this phase, seen more clearly in Fig.~\ref{fig-PE-T0-small}, can, however, exist
as a metastable state only.

%We note that at $0 < n < 0.16$~fm$^{-3}$ there appears a mixed phase

Note that at any $T>0$ the chemical potential has a well-defined limiting behavior
$\mu\to -\infty$ at $n \to 0$. 
%
%(see Appendix~\ref{app:GibbsMaxwell}). 
At  $T=0$ the situation is different:
%$\mu$ has no limiting behavior as $n \to 0$. Formally, 
at $n=0$ the chemical potential may have any value smaller than the particle mass.
The mixed  gas-liquid phase at $T=0$ 
is depicted by the horizontal lines in Fig.~\ref{fig-PE-T0} (a) and (b).
The two coexisting phases at $T=0$
are the liquid phase with $n=0.16$~fm$^{-3}$, $p=0$, 
and $\mu = 922$~MeV, 
and the gaseous phase with $n=0$, $p=0$ 
and $\mu = 922$~MeV.
This corresponds to 
%One can see explicitly that these two phases fulfill 
the Gibbs conditions of phase equilibrium, 
i.e., equal temperatures, pressures, and chemical potentials for coexisting phases.
%i.e. that $T_g = T_l$, $P_g=P_l$, and $\mu_g = \mu_l$, however, one sees that 
The stable gaseous phase at $T=0$ is, in fact, a vacuum with $n=0$.

\subsection{Phase diagram}

\begin{figure}[t]
\centering
\includegraphics[width=0.49\textwidth]{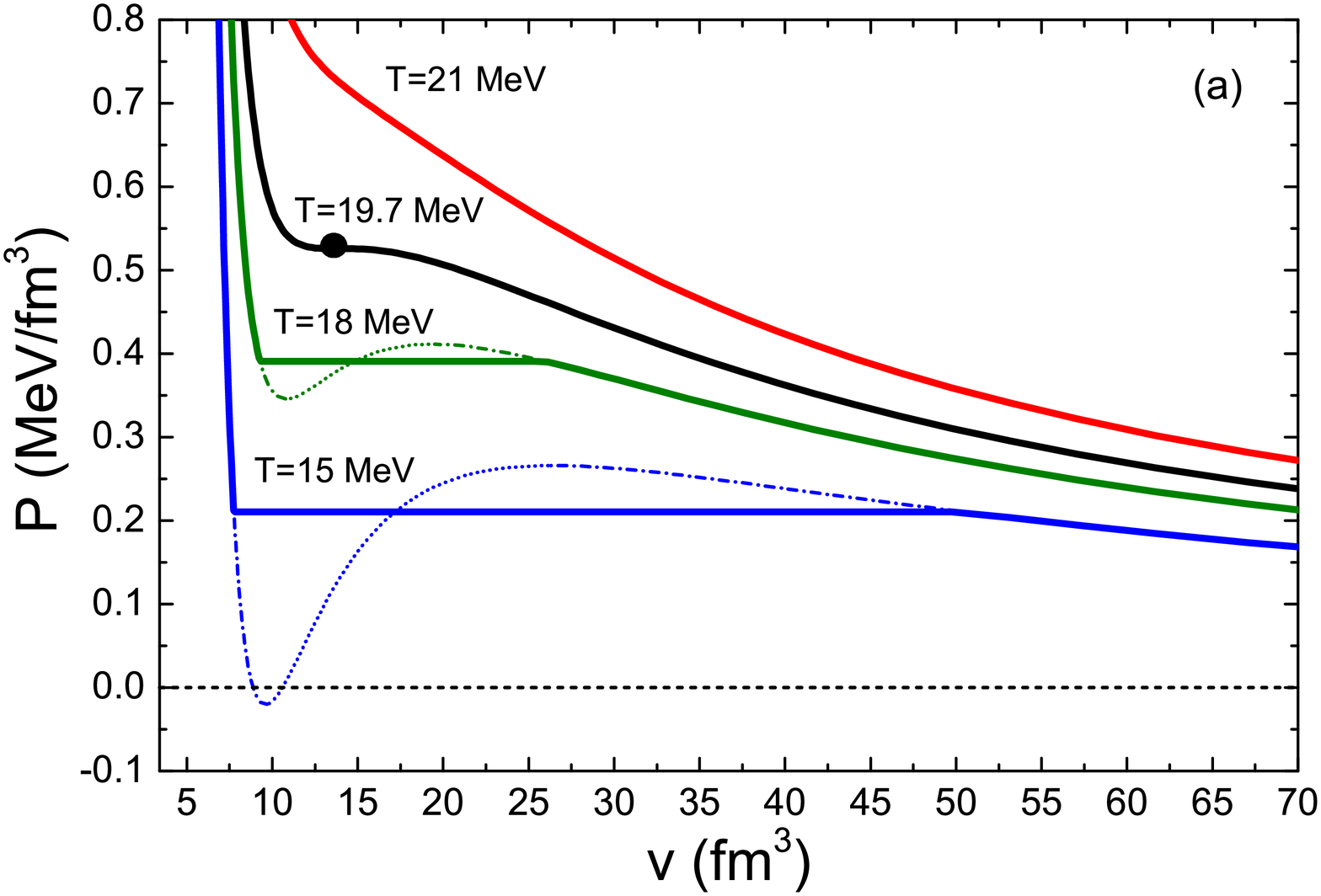}
\includegraphics[width=0.49\textwidth]{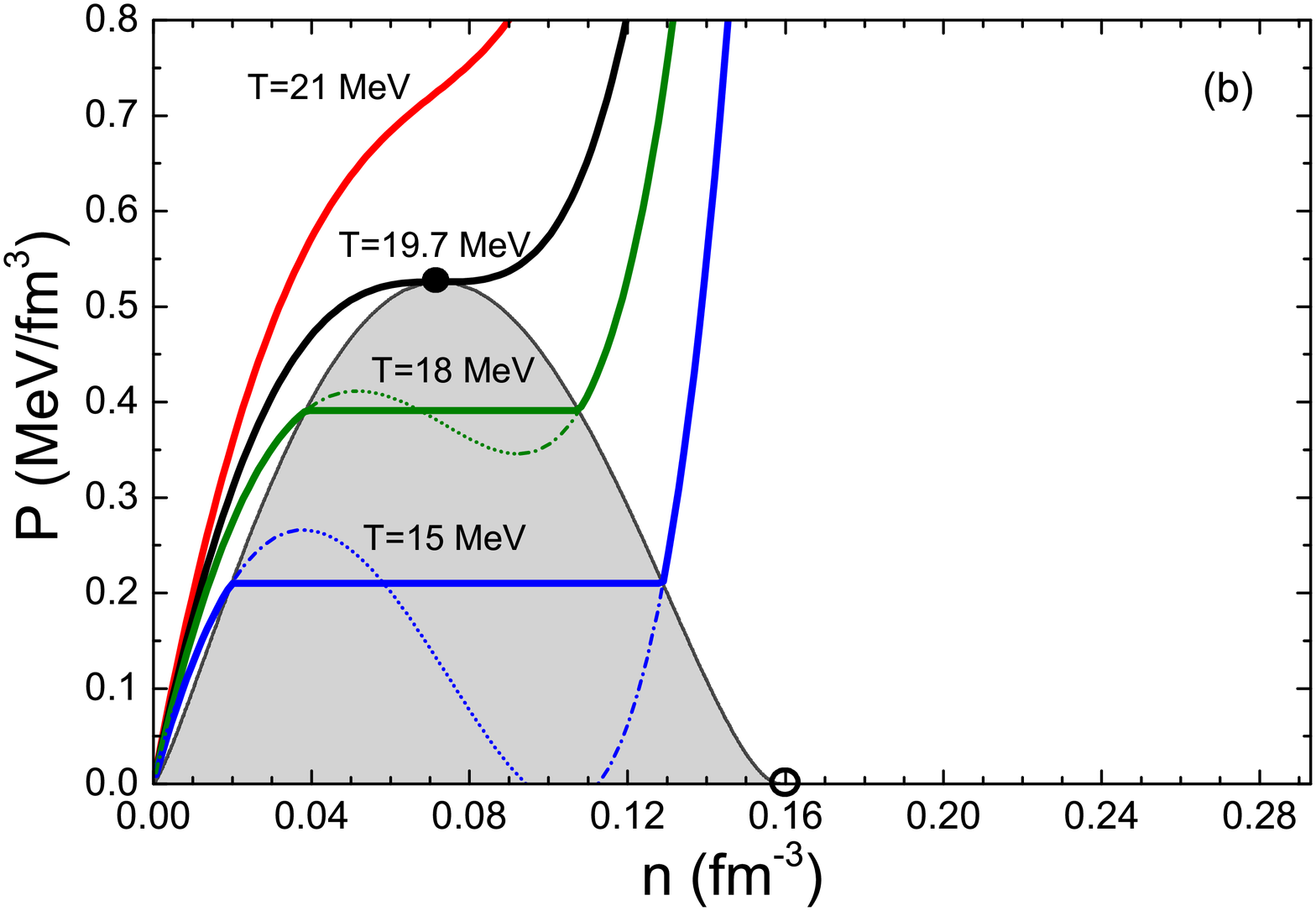}
\caption[]{Pressure isotherms in (a) $(p,v)$
and (b) $(p,n)$ coordinates, calculated in the quantum
Van der Waals equation of state with
parameters $a \cong 329$~MeV$\cdot$fm$^3$ and
$b \cong 3.42$~fm$^3$ ($r \cong 0.59$~fm).
The dashed-dotted lines
present the metastable parts of the VDW
isotherms at $T<T_c$, whereas
the dotted lines correspond to unstable parts.  The full circle
on the $T=T_c$ isotherm corresponds to the critical point, while the open circle
at $T=0$ in (b) shows the ground state of nuclear matter.
Shaded grey area in (b) depicts the mixed phase region
obtained from the Maxwell construction of equal areas
for $p(v)$ isotherms in (a).
}\label{fig-Pvn}
\end{figure}

The VDW pressure isotherms are
depicted
in $(T,v)$ and $(T,n)$ coordinates ($v\equiv 1/n$) in Fig.~\ref{fig-Pvn}
(a) and (b), respectively. They are calculated within the quantum
VDW equation of state using Eq.~\eqref{eq:PnTq}
with
$a \cong 329$~MeV$\,$fm$^3$ and
$b \cong 3.42$~fm$^3$.
The critical temperature is found to be $T_c \cong 19.7$~MeV.
%As was mentioned, the liquid-gas phase
%transition was observed experimentally.
The value of the critical temperature in our model 
is close to the experimental
estimates in Refs.~\cite{Natowitz,Karnaukhov}.
At $T<T_c$
two phases
appear: the gas and liquid phases separated by a first-order phase transition.
The mixed phase region is obtained from the Maxwell construction of equal areas for $p(v)$ isotherms
(see Appendix~\ref{app:GibbsMaxwell}),
and it is depicted by horizontal lines in Fig.~\ref{fig-Pvn} (a)
and by the shaded grey area in Fig.~\ref{fig-Pvn} (b).
%The horizontal lines correspond to the Maxwell construction of equal areas
%(see Appendix~\ref{app:GibbsMaxwell}) for $p(v)$ isotherms in Fig.~\ref{fig-Pvn} (a), 
%and the shaded grey area in Fig.~\ref{fig-Pvn}
%(b) depicts
%the resulting mixed phase region.
%
The nucleon number density at the critical point is found to be
$n_c \cong 0.07$~fm$^{-3} \cong 0.4 \, n_0$. Normal nuclear matter with $n=n_0\cong 0.16$~fm$^{-3}$
and $T=0$ corresponds to a point placed exactly
on the boundary between the mixed and liquid phases.
Note also that the maximal value of the nucleon number density
in the VDW model is $n_{\rm max}=1/b$, which is equal to $n_{\rm max} \cong 0.29$~fm$^{-3}$
for the chosen value of parameter $b$.

In the mixed phase region, the particle number density is given by
\eq{\label{n-mix}
n~=~\xi n_g~+~(1-\xi) n_l~,
}
where $\xi$ and $1-\xi$ are the volume fractions of the gaseous
and liquid components, respectively.
The values of $n_g$ and $n_l$ in Eq.~(\ref{n-mix})
are the particle densities of, respectively, the gaseous and liquid phases
at the corresponding boundaries with the mixed phase. A behavior of the mixed phase
at $T=0$ is rather special. The 
stable gaseous phase is absent at $T=0$, i.e., $n_g=0$
at the mixed phase boundary as this boundary starts
from the point $T=0$ and $n=0$. 
Therefore, only a metastable gaseous phase
at small densities can exist at $T=0$ as depicted in Fig.~\ref{fig-PE-T0-small}.
The stable
gaseous phase exists however at any $T>0$
for small enough values of the particle number density,
smaller than the $n_g$ density of gaseous phase in the 
mixed phase region resulting from the Maxwell construction.

Parameters of the critical point found in the VDW case with Fermi
statistics for nucleons differ significantly from those values
for the classical VDW gas.   With
the same VDW parameters $a$ and $b$ as in the Fermi statistics, the
classical VDW equation (\ref{eq:vdw}), i.e., with Boltzmann statistics, would give
$T_c = 8a/27b \cong 28.5$~MeV and
$n_c = 1/3b \cong 0.10$~fm$^{-3}$. This
further
indicates an importance of the effects of quantum statistics:
these effects are not only crucial in the limit $T\rightarrow 0$
but also remain quantitatively important even near the critical point.

\begin{figure}[ht]
\centering
\includegraphics[width=0.89\textwidth]{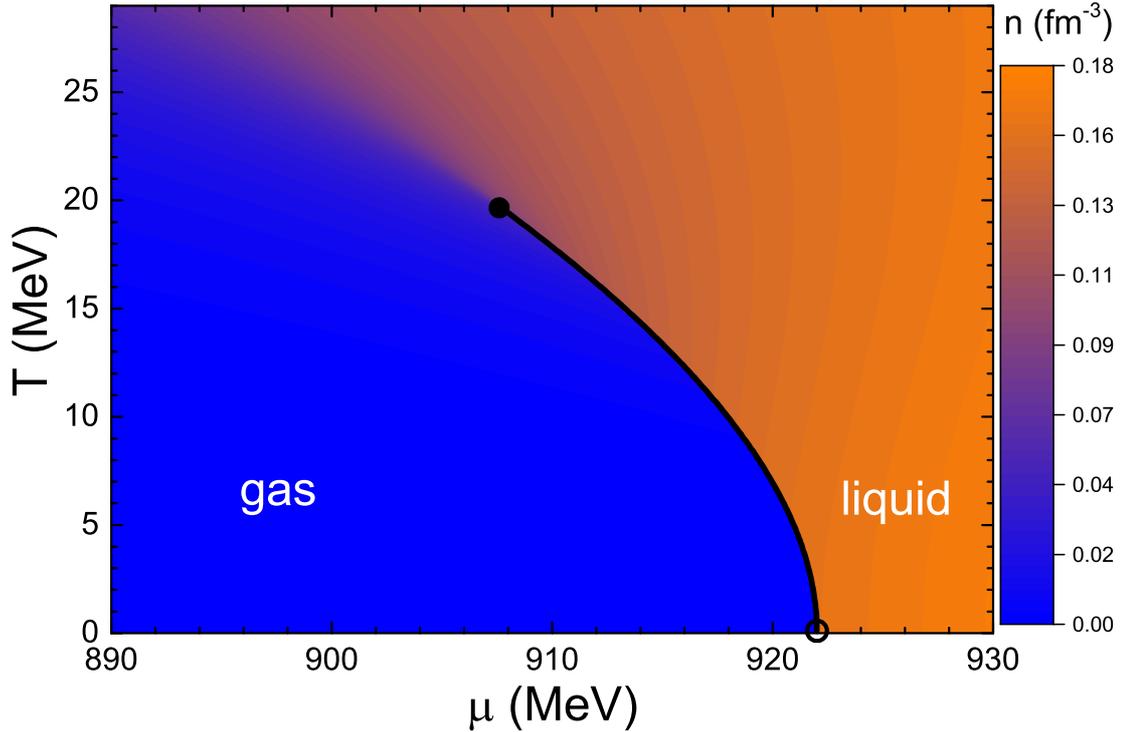}
\caption[]{Density of the symmetric nuclear matter
in $(T,\mu)$ coordinates, calculated in the quantum
van der Waals equation of state with
parameters $a \cong 329$~MeV$\,$fm$^3$ and
$b \cong 3.42$~fm$^3$ ($r \cong 0.59$~fm).
The open circle denotes the ground state of nuclear matter,
the full circle corresponds to the critical endpoint,
and the phase transition curve is depicted by the solid black line.
}\label{fig-Tmu-n}
\end{figure}

In Fig.~\ref{fig-Tmu-n} the phase diagram of the symmetric nuclear matter
in $(T,\mu)$ coordinates is depicted. The nucleon density at different
temperature $T$ and chemical potential $\mu$ is presented.
At $T<T_c$ there is the $T$-$\mu$ region
with three different solutions for $p(T,\mu)$
at given $T$ and $\mu$. According to the Gibbs criterion, a solution
with a largest pressure survives (see Appendix~\ref{app:GibbsMaxwell}), and only this solution for $n(T,\mu)$  is depicted
in Fig.~\ref{fig-Tmu-n}.
The location of the critical point is shown
in Fig.~\ref{fig-Tmu-n} by the
solid circle, while the normal nuclear matter state
corresponds to the open circle. Note that $\mu_0\cong 922$~MeV
corresponds to the chemical potential of the normal nuclear matter which is
placed on a boundary with the liquid phase.
The values $\mu<\mu_0$ are forbidden at $T=0$ (these values of $\mu$
lead formally to $n=0$). The values $\mu>\mu_0$ at $T=0$ are possible
and correspond to the nuclear liquid.

The phase transition line, $\mu=\mu_{\rm mix}(T)$, shown in Fig.~\ref{fig-Tmu-n}, starts from
the normal nuclear matter state
with $T=0,~\mu\cong 922$~MeV and ends at the critical point with $T_c\cong 19.7$~MeV,
$\mu_{\rm mix}(T_c)\cong 908$~MeV.
This line presents the whole mixed phase region shown by the grey area
in Fig.~\ref{fig-Pvn}. At each $T<T_c$ two solutions, $n_g(T,\mu)$ and $n_l(T,\mu)$,
with different particle densities, $n_g(T,\mu)<n_l(T,\mu)$, and
equal pressures, $p_g(T,\mu)=p_l(T,\mu)$, exist at the phase transition line $\mu=\mu_c(T)$.
On this line, the discontinuities of thermodynamical quantities $n$, $\varepsilon$, and $s$ take place.

At $T>T_c$ there is only one solution $n(T,\mu)$  for any $T$ and $\mu$ values,
i.e., there are no distinct {\it gaseous} or {\it liquid} phases.
Nevertheless, as seen from Fig.~\ref{fig-Pvn},
very rapid, although continuous, changes of particle number density take place in a narrow
$T$-$\mu$ region even at $T>T_c$. This is a manifestation of the so-called smooth crossover phenomenon.

At any $T>0$,
there are no restrictions on possible values of the chemical potential;
i.e., any values of $\mu$ between $-\infty$ and $+\infty$ are possible.  When $\mu$ decreases
the particle number density decreases too and goes to zero at $\mu\rightarrow -\infty$.
At very small $n$, both the particle interactions and Fermi statistics effects become negligible.
The system of nucleons  behaves then as the ideal Boltzmann gas. In an opposite limit,
$\mu\rightarrow\infty$,  nucleon density $n(T,\mu)$ goes to its upper limiting value $1/b$.
The VDW pressure behaves then approximately as $p\cong nT/(1-bn)$ and goes to infinity.
Different theoretical models and their comparison
with experimental estimates of the nuclear matter properties
have been widely discussed in the literature (see, e.g.,
Refs.~\cite{Blaizot,Shlomo,Stone:2014wza}).
In the present paper we do not attempt to make any
detail comparison of the developed VDW quantum  model with
existing data for nuclear matter.
Some extensions of the model
will probably be needed.
These questions are, however, beyond the scope of the present paper.

\section{Summary}
\label{sec:summary}

In the present paper we have formulated a generalization
of the van der Waals equation of state to include effects
of quantum statistics. In the grand canonical ensemble a system of
two transcendental equations
for the pressure
and particle density is obtained.
These equations can be solved
for all possible values of temperature, $T\ge 0$, and chemical potential,
$-\infty <\mu<\infty$.
Our quantum generalization of the VDW equation satisfies all basic
requirements:
it reduces to the ideal Fermi or Bose gas for $a=0$ and $b=0$,
to the classical VDW equation in the Boltzmann limit, and it
satisfies the 3rd law of thermodynamics, i.e. $s\to0$ as $T \to 0$.

The VDW equation with Fermi statistics has then been applied
to a system of interacting nucleons
to describe the properties
of symmetric nuclear matter. The VDW parameters $a$ and $b$ of
interacting nucleons are fixed
by the properties of the nuclear matter ground state: $T=0$, $p=0$,
$n=n_0=0.16$~fm$^{-3}$, and $E_B=-16$~MeV. We find $a \cong 329$~MeV$\,$fm$^{3}$
and $b \cong 3.42$~fm$^{3}$.
With these parameters the VDW model predicts
a first-order liquid-gas phase transition with a critical endpoint located
at $T_c \cong 19.7$~MeV and $n_c \cong 0.07$~fm$^{-3}$.
Extensions
of the presented formulation as well as new physical applications
will be the subject of further studies.

\begin{acknowledgments}
We would like to thank M. Ga\'zdzicki, A.~G.~Magner, S. Mr\'owczy\'nski, and K. Redlich
for fruitful comments and discussions.
This work was partially supported
by HIC for FAIR within the LOEWE program of the State of Hesse
and by the Program of
Fundamental Research of the Department of Physics and Astronomy of National Academy of Sciences of Ukraine.
\end{acknowledgments}

%%%%%%%%%%%%%%%%%%%%%%%%%%%%%%%%%%%%%%%%%%%%%%%
\appendix
\section{Gibbs criteria and  Maxwell construction}
\label{app:GibbsMaxwell}
\numberwithin{equation}{section}
%%%%%%%%%%%%%%%%%%%%%%%%%%%%%%%%%%%%%%%%%%%%%%%
A statistical system in the GCE is defined by two independent variables, $T$ and $\mu$.
Two distinct phases, {\it gas} and {\it liquid}, coexist if their pressures are equal, $p_g(T,\mu)=p_l(T,\mu)$.
In the case of $p_g(T,\mu)\neq p_l(T,\mu)$, only a phase with a larger pressure survives.
These statements are known as the Gibbs
criteria for the the first-order phase transition (see, e.g., Refs.~\cite{greiner,LL}).
We now prove that the Gibbs criteria are equivalent to the Maxwell construction of equal areas for the VDW equation of state.

\begin{figure}[ht]
\centering
\includegraphics[width=0.49\textwidth]{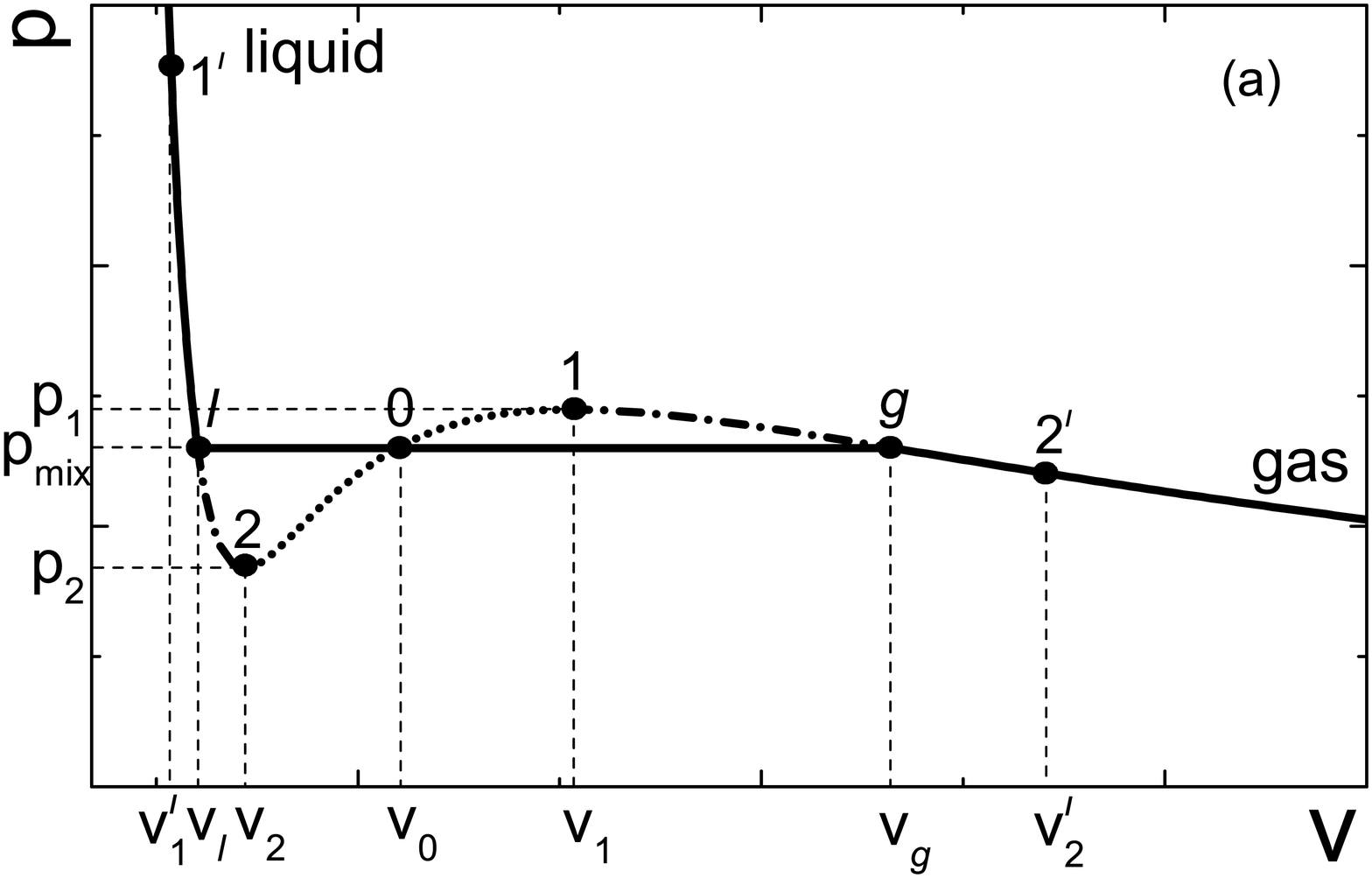}
\includegraphics[width=0.49\textwidth]{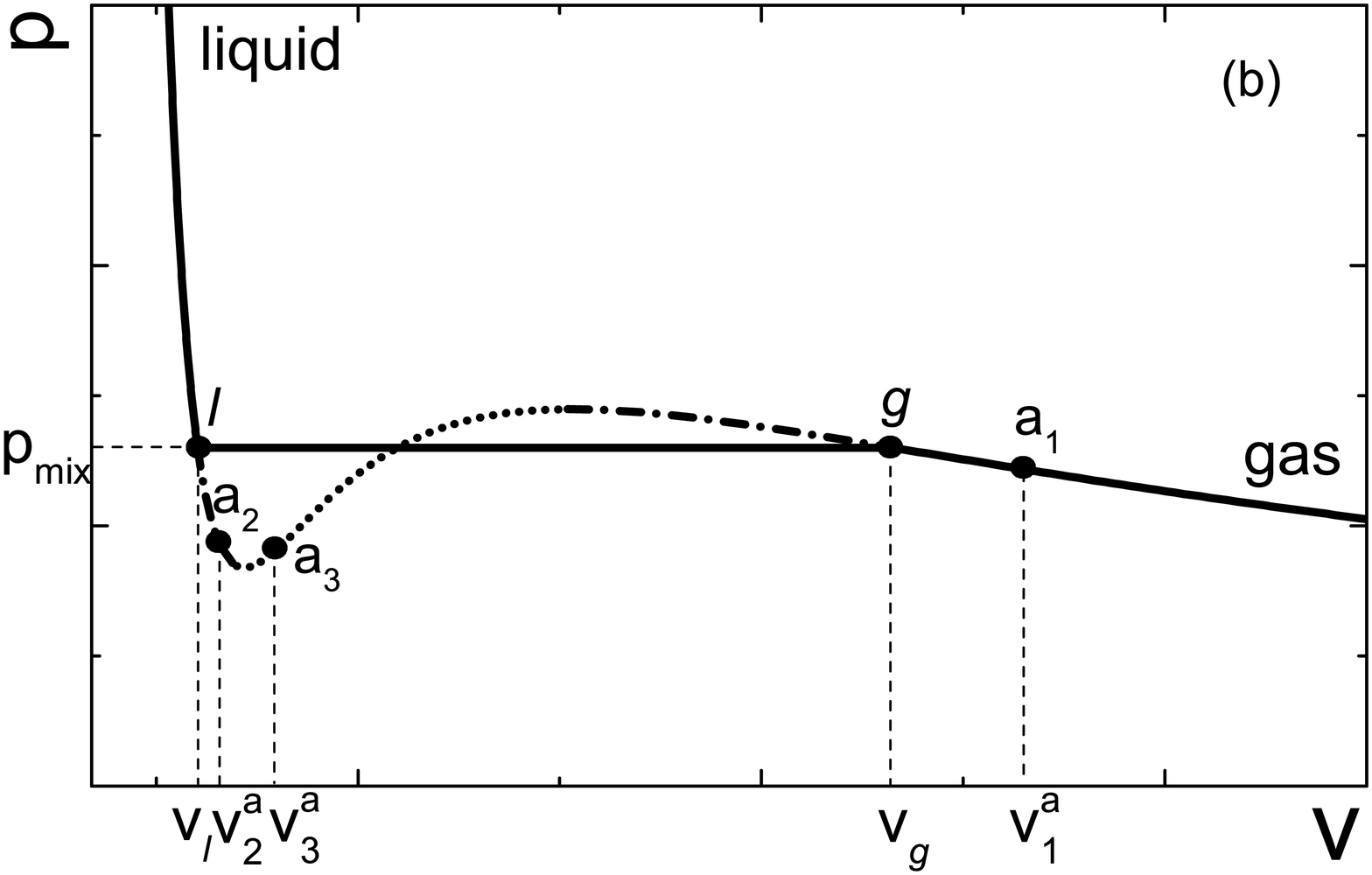}%
\caption[]{
%{\bf
A generic VDW isotherm $p(v,T)$ for $T<T_c$.
Points $l$ and $g$ on the isotherm correspond to the boundaries of the mixed phase.
In (a) point 1 (2) on the isotherm corresponds to the maximum (minimum)
value of the chemical potential inside the mixed phase, which is reached at the boundary between metastable and unstable phases, while point $1'$ ($2'$)
corresponds to the same value of the chemical potential reached in the pure liquid (gaseous) phase. In (b) points $a_1$ (gaseous phase), $a_2$ (metastable liquid phase) and $a_3$ (unstable phase) which all correspond to the same value of the chemical potential are depicted on the VDW isotherm.
%}
}\label{fig-MG}
\end{figure}

At $T<T_c$, the Maxwell construction replaces a part of the VDW isotherm $p=p(v,T)$
by the horizontal line $p=p_{\rm mix}$, which corresponds to the mixed phase region for all $v$ in the interval
$[v_l,v_g]$. This is shown in Fig.~\ref{fig-MG}. The Maxwell equal areas are
\eq{\label{Max}
\int_{v_l}^{v_0}dv\, \left[p_{\rm mix}~-~p(v,T)\right]~=~\int_{v_0}^{v_g}dv\, \left[p(v,T)~-~p_{\rm mix}\right]~,
}
where $v_l<v_0<v_g$ and $p(v_0,T)=p_{\rm mix}$.
The replaced parts of the isotherm are interpreted
as metastable ($\partial p / \partial v < 0$) and unstable
($\partial p / \partial v > 0$) states. They are shown in Fig.~\ref{fig-MG}
by the dashed-dotted and dotted lines, respectively.

Using the thermodynamical identity
\eq{
%\left(\frac{\partial G}{\partial p}\right)_{N,T} =
\left(\frac{\partial \mu}{\partial p}\right)_{T} ~=~ v~,
}
one can present the chemical
potential $\mu_A$ at any point $A$ on the isotherm as
\eq{\label{eq:mupath}
\mu_A ~=~ \mu_B ~+~ \int_{p_B}^{p_A} \, dp' \, v(p',T)~,
}
where $B$ is an arbitrary point on the isotherm, and
integration in (\ref{eq:mupath}) is performed along the path from point
$B$ to point $A$ on the isotherm.
With Eqs.~(\ref{Max}) and (\ref{eq:mupath}) one can easily prove that
\eq{\label{mu-mix}
\mu(v_l,T)~=~\mu(v_g,T)~\equiv ~\mu_{\rm mix}~,
}
thus, the Maxwell and Gibbs constructions for the mixed phase are equivalent.

According to Eq.~(\ref{eq:mupath}) the chemical potential decreases with $v$ if $(\partial p/\partial v)_T <0$, and
increases if $(\partial p/\partial v)_T >0$. Therefore, inside the mixed phase region $[v_l,v_g]$ the chemical potential
reaches its minimal value $\mu=\mu_2<\mu_{\rm mix}$ at $v=v_2$  and its maximal value $\mu=\mu_1>\mu_{\rm mix}$ at $v=v_1$.
The points $v_1$ and $v_2$, where $(\partial p/\partial v)_T=0$, correspond to the boundaries between the metastable and
unstable parts of the VDW isotherm.
At $v<v_l$ and $v>v_g$ the chemical potential is a monotonously decreasing function of $v$, with $\mu\rightarrow \infty$
at $v\rightarrow b$ and $\mu\rightarrow - \infty$
at $v\rightarrow \infty$. Therefore, there is a point $v_2'>v_g$ in the gaseous phase where $\mu=\mu_2$,
and a point $v_1'<v_l$ in the liquid phase where $\mu=\mu_1$. These two points are depicted
in Fig.~\ref{fig-MG} (a). At both $\mu>\mu_1$ and $\mu<\mu_2$ the GCE VDW pressure $p(T,\mu)$ is a unique function.
On the other hand, there are three different solutions for the VDW pressure at $\mu_2<\mu<\mu_1$.

Let us first consider $\mu_2<\mu_a<\mu_{\rm mix}$. There are three points on the VDW isotherm
with $\mu=\mu_a$ shown in Fig.~\ref{fig-MG} (b): point $a_1$ in the gaseous phase with $v_g<v^a_{1}<v_2'$,
point $a_2$ in the metastable liquid phase
with with $v_l<v^a_{2}<v_2$, and  point $a_3$ in the unstable phase with $v_2<v^a_{3}<v_1$.
Using Eq.~(\ref{eq:mupath}) one finds that
\eq{\label{a1-a2}
\int_{p^a_1}^{p_{\rm mix}}dp'\,v(p',T)~=~\int_{p^a_2}^{p_{\rm mix}}dp'\,v(p',T)~ =~
\int_{p_2}^{p_{\rm mix}}dp'\,v(p',T)~-~\int_{p_2}^{p^a_3}dp'\,v(p',T)~.
}
Applying the mean value theorem to integrals in Eq.~(\ref{a1-a2}) one
obtains
\eq{\label{pa1-pa2}
 \left(p_{\rm mix}~-~p^a_1\right)\,\overline{v}_1~=~\left(p_{\rm mix}~-~p^a_2\right)\,\overline{v}_2~=~
\left(p_{\rm mix}~-~p_2\right)\,\overline{v}_{31}
-~\left(p_{2}~-~p^a_3\right)\,\overline{v}_{32}~,
}
where $v_l<\overline{v}_2<v^a_2$,  $v_g<\overline{v}_1<v^a_1$,
$v_l<\overline{v}_{31}<v_2<\overline{v}_{32}<v^a_3$, and $p_2=p(v_2,T)$.
It follows from Eq.~(\ref{pa1-pa2})
that $p^a_1>p^a_2$ and $p^a_1>p^a_3$; i.e., at $\mu_2<\mu<\mu_{\rm mix}$ the gaseous phase
should be realized according to the Gibbs criterium as its pressure is larger
than pressures of both metastable and unstable states.

The same arguments are applied to $\mu_{\rm mix}<\mu<\mu_{1}$ and show that the liquid pressure
is then larger
than the pressures of both metastable and unstable states.
Therefore, for the VDW equation of state the Maxwell construction of the equal areas
and Gibbs criteria are fully equivalent. The Maxwell construction is applied
in the CE, whereas the Gibbs criteria are used in the GCE.
We emphasize that this statement is valid not only for the
classical VDW equation of state, but also for 
the VDW equation
with Fermi statistics with isotherms depicted in Fig.~\ref{fig-Pvn}.
%{\bf
%any of the $T>0$ isotherms of
%}
%the VDW equation
%with Fermi statistics depicted in Fig.~\ref{fig-Pvn}.

\section{Thermodynamic mean-field approach}
\label{app:MF}
\numberwithin{equation}{section}
In the framework of the thermodynamic mean-field (TMF) approach~\cite{mf-1992,mf-1995,mf-2014} 
the pressure and  particle number density are presented as
\begin{eqnarray}
p(T,\mu) & ~=~ & p^{\rm id} [T, \mu_{\rm id} (n,T)] ~+~ P^{\rm ex} (n,T)~,
\label{eq:PMF} \\
n(T,\mu) & ~=~ & n^{\rm id} [T, \mu - U(n,T)]~,
\label{eq:UMF}
\end{eqnarray}
where $P^{\rm ex}(n,T)$ and $U(n,T)$ are, respectively, the excess pressure and the thermodynamic mean field. 
The presence of nonzero quantities $P^{\rm ex}(n,T)$ and $U(n,T)$ in Eqs.~(\ref{eq:PMF}) and (\ref{eq:UMF})
correspond to interaction between particles, and the condition of thermodynamic consistency reads \cite{mf-2014}:
\eq{\label{eq:UPMF}
n \frac{\partial U}{\partial n}~ = ~\frac{\partial P^{\rm ex}}{\partial n}~.
}
For a specific choice of $P^{\rm ex}(n,T)$ and $U(n,T)$ functions,
one proceeds by solving of Eq.~\eqref{eq:UMF} for $n=n(T,\mu)$, 
and then the pressure $p(T,\mu)$ can be obtained from Eq.~\eqref{eq:PMF}
(see some examples in Ref.~\cite{mf-2014}).

The VDW equation of state with quantum statistics defined by Eqs.~\eqref{eq:pq} and \eqref{eq:nvdwq} 
can be rewritten in the TMF form (\ref{eq:PMF}) and (\ref{eq:UMF}).
In order to determine $P^{\rm ex}(n,T)$ and $U(n,T)$ we rewrite Eq.~\eqref{eq:PnTq} 
for the VDW pressure as
\eq{\label{eq:PVdWq2}
p ~=~ p^{\rm id} [T,\mu^{\rm id} (n,T)] ~+~ p^{\rm id} 
\Big[T,\mu^{\rm id}\Big(\frac{n}{1-bn},T\Big)\Big] ~-~ p^{\rm id} [T,\mu^{\rm id} (n,T)] ~-~ a\,n^2~.
}
Comparing \eqref{eq:PMF} and \eqref{eq:PVdWq2} one finds
\eq{\label{eq:PMFVdWq}
P^{\rm ex}_{\rm VdW}(n,T) ~= ~p^{\rm id} \Big[T,\mu^{\rm id}\Big(\frac{n}{1-bn},T\Big)\Big] ~-~
 p^{\rm id} [T,\mu^{\rm id} (n,T)] ~-~ a\,n^2~.
}
The mean field $U_{\rm VdW}(n,T)$ can be then calculated from
Eq.~\eqref{eq:UPMF} as
\eq{\label{eq:UMFVdWq}
U_{\rm VdW}(n,T) ~= ~\int_0^n \, \frac{1}{n'} \, \frac{\partial P^{\rm ex}_{\rm VdW} (n',T)}{\partial n'} \, d n'~.
}
For the Boltzmann statistics, Eqs.~(\ref{eq:PMFVdWq}) and (\ref{eq:UMFVdWq})
are simplified to the following analytical expressions:
\eq{\label{eq:PMFVdW}
& P^{\rm ex}_{\rm VdW}(n,T) ~=~ Tn \frac{bn}{1-bn} ~-~ an^2~,\\
\qquad
& U_{\rm VdW} (n,T)~ =~ T \frac{bn}{1-bn} ~-~ T \ln(1-bn) ~-~ 2 a n~.\label{eq:UMFVdW}
}

\end{document}